\documentclass[a4paper,twocolumn,aps,floatfix,twoside,showpacs]{revtex4}

\usepackage{graphicx}
\usepackage{bm}
\usepackage{amstext} 
\usepackage{amssymb} 

\begin{document}

\title{
Does frequency-temperature superposition hold in deeply super-cooled liquids?
}

\author{Catalin Gainaru}
\author{Alexander Brodin}
\author{Vladimir Novikov}
\altaffiliation[Permanent address: ]{Institute of Automation and Electrometry,
	Novosibirsk 630090, Russia}
\author{Ernst A. R\"ossler}
\affiliation{Experimentalphysik II, Universit\"at Bayreuth, D-95440 Bayreuth, Germany}

\date{\today}

\begin{abstract}
The temperature evolution of the broadband $10^{-6}$-$10^{10}$ Hz dielectric susceptibility of the paradigmatic glass formers glycerol, propylene carbonate, and fluoro-aniline is analyzed assuming a three-step relaxation due to the $\alpha$-process, its excess wing, and a $\beta$-process. We find that the $\alpha$-peak and the wing can be described by susceptibility functions with temperature-independent high-frequency exponents, while the relative weight of these contributions does depend on the temperature. The excess wing and the $\beta$-process are distinct phenomena; in particular, the relaxation strength of the excess wing grows with decreasing the temperature, contrary to that of the $\beta$-process. In our interpretation, the frequency-temperature superposition of the $\alpha$-process is valid for all temperatures; in the case of glycerol, a typical $\beta$-process is unambiguously identified for the first time.
\end{abstract}


\maketitle

Understanding the evolution of the dynamic susceptibility in super-cooled molecular liquids is one of the central topics currently addressed by many scientists. Since the early days of glass research, the frequency-temperature superposition (FTS), i.e. the near independence of the shape of the primary non-exponential relaxation ($\alpha$-process) of temperature, is considered an important property of the glass transition phenomenon. Experimental progress in dielectric spectroscopy revealed, however, deviations that seem to increase on approaching the glass transition temperature $T_g$. The situation is further obscured by the emergence, close to $T_g$, of secondary relaxation features, notably the so-called excess wing seen as an extra power-law wing on the high-frequency side of the $\alpha$-peak. 

As there is no theory to address these dynamics, the attempts to describe the co-evolution of the $\alpha$-process and excess wing are necessarily based on phenomenological grounds. Dixon {\it et al.} \cite{Dixon90} proposed a scaling of the dielectric loss yielding a master curve for both the $\alpha$-peak and excess wing. Their scaling analysis suggests that the exponents characterizing the two apparent power-laws of the susceptibility on the high frequency side of the relaxation peak are related \cite{Dixon90}. While the exactness of the scaling may be arguable \cite{Kudlik90}, it still works remarkably well in many cases \cite{Lunk2000}. Building upon this approach, even a divergence of the static susceptibility below $T_g$ was suggested \cite{Menon95}. In another approach, the slow dynamics is decomposed into two separate relaxation processes \cite{Lunk2000}. This is partly corroborated by recent aging experiments of Schneider {\it et al.} \cite{Schneider2000} that show the secondary relaxation features becoming more pronounced after aging. The authors concluded that the excess wing is just a secondary relaxation process similar to the $\beta$-relaxation discussed by Johari and Goldstein \cite{Johari70} and found in many systems. Some glassformers exhibit, however, both a well-resolved $\beta$-peak {\it and} excess wing \cite{Kudlik90,Blo03,Ngai2004,Adic03}. Thus, it remains unclear whether the excess wing and $\beta$-relaxation are similar in nature, or rather they are distinct phenomena.

Recently, Blochowicz {\it et al.} \cite{Blo03,Blo06} analyzed dielectric spectra of several glassformers that did not show a discernible $\beta$-peak and obtained perfect fits, using a model based on an extension of the generalized gamma distribution of relaxation times. They found high correlation between the distribution parameters as obtained from fits of many glassformers. Furthermore, above a temperature $T_x$ their fits tended to converge to simple Cole-Davidson (CD) spectra with temperature-independent widths, thus complying with FTS at $T>T_x$. The crossover temperatures $T_x$ appeared to coincide with the so-called Stickel temperatures of the same materials \cite{Blo06}, which mark a crossover from the low-temperature Vogel-Fulcher-Tammann behavior to a different temperature dependence of the relaxation times. The unmistakable correlation between the shape parameters of the $\alpha$-peak and excess wing at $T<T_x$ was interpreted to imply a close connection between the processes.

Though different in detail, all the mentioned approaches agree that the width parameters of the $\alpha$-peak and wing both vary with temperature. Similar findings were reported for the $\alpha$-peak width of plastically crystalline systems that do not show any excess wing \cite{Gainaru06}. Thus, at least at low temperatures, FTS appears not to apply. This may have far reaching consequences for any theoretical description. We remind the reader that within the mode coupling theory, being the most advanced approach to understanding the onset of the glass transition at high temperatures, the applicability of FTS for the stretched long-time end of the correlation functions is an essential prediction that is well confirmed by light and neutron scattering experiments at high temperatures \cite{Petry95,Cummins97,Brodin05}.

\begin{figure}
\includegraphics[width=.9\columnwidth]{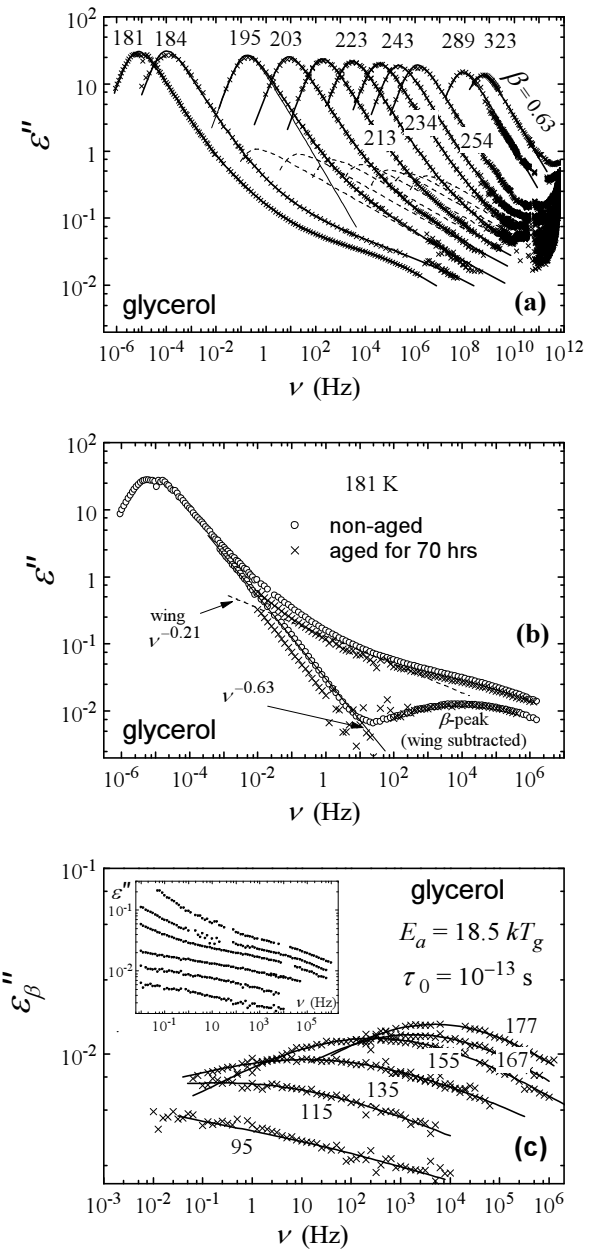}
\caption{\label{glycerol} {\bf (a)} Dielectric loss of glycerol at indicated temperatures with fits. Dashed lines are excess wing contributions.  {\bf (b)} Aging experiment and its analysis, see text for details. 
{\bf (c)} $\beta$-peaks of glycerol, extracted from the dielectric spectra shown in the inset, with fits.
	}
\end{figure}

In this paper we suggest, at variance with current approaches, that the $\alpha$-peak and excess wing indeed can be described by susceptibility functions with {\it temperature-independent} high-frequency exponents in the {\it whole} temperatures range down to $T_g$ and, with reservations, even below. In order to understand the aging experiments of Schneider {\it et al.} \cite{Schneider2000} and ours (see Fig. \ref{glycerol}(b) and discussion), as well as to account for a negative curvature at the high-frequency end of our extended data of glycerol at $T \gtrsim T_g$ (see $T=181$ and 184 K in Fig. \ref{glycerol}(a) and discussion), we allow for the presence of a $\beta$-peak in addition to the $\alpha$-peak and excess wing. In a sense, we extend the approach of Blochowicz {\it et al.} \cite{Blo03,Blo06} and allow for a $\beta$-process in cases where they did not, as none was clearly discernible in their more limited data. Though this implies more parameters, we are able to keep most of them fixed and obtain equally good fits as in Ref. \cite{Blo03} with at least {\it one less} free parameter (see later for the enumeration of the parameters). At the same time, we are able to recover FTS and, in addition, to single out a typical $\beta$-process. We note that recently Olsen {\it et al.} \cite{Olsen01} renewed the discussion of the applicability of FTS. Disregarding any influence from secondary processes, they claimed a universal high-frequency exponent $\beta = 0.5$ at $T_g$. Of course, this must imply a substantial temperature dependence of the $\alpha$-peak width, since at the highest temperatures one usually finds a much larger stretching parameter.

In order to demonstrate our analysis, we recourse to the best dielectric data currently available. Thus, we display in Fig. \ref{glycerol}(a) the dielectric loss of glycerol due to Lunkenheimer et al. \cite{Lunk2000} together with some previously \cite{Blo03} and newly obtained data from our group. One clearly recognizes the emergence of the excess wing on approaching $T_g$, while, in contrast, no indication of a wing can be seen above say 250 K. On a closer inspection of the excess wing at 181 and 184 K, an indication of a crossover from a positive to negative curvature is observed, which we take as an indication that a $\beta$-peak may be involved. With this in mind and in contrast to existing approaches, we assume that {\it three} contributions, an $\alpha$-peak, excess wing, and $\beta$-peak, are present. Then, the {\it apparent} excess wing includes the contribution of a weak $\beta$-process that obscures the actual shape of the wing.

\begin{figure}
\includegraphics[width=.9\columnwidth]{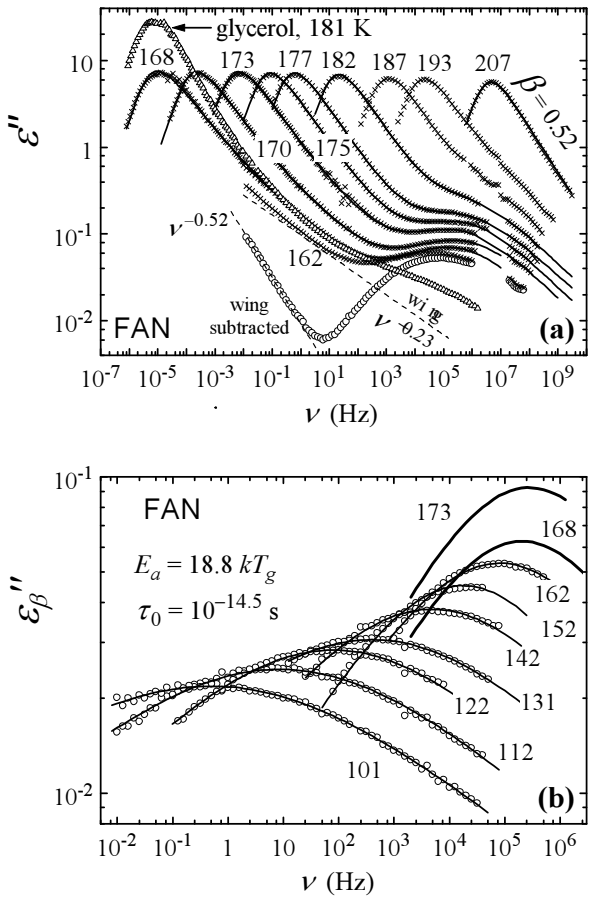}
\caption{\label{FAN} {\bf (a)} Dielectric loss of FAN at indicated temperatures with fits; the lowermost dataset (circles) is the 162 K dataset with the excess wing subtracted out.
{\bf (b)} $\beta$-peaks of FAN, extracted from its dielectric data at $T \leq 162$ K (not shown), with fits. Data at 168 and 173 K are generated from the fit functions.
	}
\end{figure}

To quantitatively describe the data, we assume a three-step relaxation function and express it as a product of three terms, $F(t)= F_\beta (t) F_{ex}(t) \phi_\alpha(t)$, where $F_\beta(t)$ and $F_{ex}(t)$ denote the partial relaxation due to the $\beta$-process and excess wing, respectively, and $\phi_\alpha(t)$ the main relaxation due to the $\alpha$-process. Introducing the normalized ($\phi(0)=1$) relaxation functions $\phi_i(t)$ and their strengths $S_i$, we write
\begin{equation}
F(t)= \left[ (1-S_\beta)\phi_\beta(t)+S_\beta \right]
	\left[(1-S_{ex})\phi_{ex}(t)+ S_{ex}\right]
	\phi_\alpha(t)
\label{rel_fn}
\end{equation}
Thus, we formally describe the excess wing and $\beta$-process as kinds of secondary processes that each relaxes, in series, a certain fraction of the total polarization. The dielectric spectrum is obtained from $F(t)$ via
$
\Delta \epsilon \, \omega \, \text{Re} \left[ {\cal F} \left( F(t) \right) \right],
$
where $\Delta \epsilon = \epsilon _0 - \epsilon _\infty$, and $\cal F$ is the Fourier transform. The $\alpha$-peak is usually well described by a simple CD function \cite{Lunk2000,Blo03,Blo06}, while the power-law spectrum of the excess wing can be equally well described by the high-frequency power-law asymptote of an additional CD function. We thus model the $\alpha$-peak and excess wing relaxation functions $\phi_\alpha(t)$ and $\phi_{ex}(t)$ by the time-domain equivalent of the CD function (the incomplete gamma function) with exponents $\beta$ and $\gamma$, respectively, and equal time constants $\tau_\alpha = \tau_{ex}$. The resulting fits are nearly perfect (see Figs. \ref{glycerol} and \ref{FAN}, and discussion), which supports our choice of the model functions. (We also tried the Kohlrausch-Williams-Watts stretched exponential function $\exp(-t/\tau)^\beta$, but could not obtain nearly as good fits.) To model $\phi_\beta(t)$ of the $\beta$-process, we use a symmetric distribution of relaxation times, suggested in Ref. \cite{Blo03}. Thus, the formal parameters of our model are: $\tau_\alpha$ and $\beta$ of $\phi_\alpha(t)$, $\gamma$ and $S_{ex}$ of $\phi_{ex}(t)$, and $S_\beta$, $\tau_\beta$, and a shape parameter $a$ of $\phi_\beta(t)$. The overall amplitude $\Delta\epsilon$ could in principle be extracted from the $\epsilon'(\omega)$ experimental data, but we found the latter not sufficiently accurate for such a purpose, and let in most cases $\Delta\epsilon$ vary in the fits.

We first analyzed the highest temperatures $\geq 289$ K of glycerol ($T_g \approx 186$ K) and found that the $\alpha$-peak is well fitted by a simple CD function ($S_\beta = S_{ex} = 1$ in Eq. (1)) with $\beta = 0.63$, see Fig. \ref{glycerol}(a). At intermediate temperatures $\geq 195$ K, we allowed for a wing, still assuming the $\beta$-contribution negligible ($1-S_\beta =0$), and obtained from unconstrained fits similar values of the $\alpha$-peak exponent $\beta \approx 0.63$. Thus, it appears that, indeed, the width of the $\alpha$-peak may be taken as temperature independent. We then assumed that the high-temperature value $\beta = 0.63$ is appropriate for {\it all} temperatures, and kept it fixed accordingly. This still leaves some freedom in the choice of the wing exponent $\gamma$, especially at the lowest temperatures, where the contribution of the $\beta$-process has to be accounted for. In fact, an accurate estimate of $\gamma$ is crucial for the analysis of the (weak) $\beta$-process. Consequently, we choose the low-temperature $\gamma$ such as to yield a {\it symmetric} $\beta$-peak (a usual property of $\beta$-processes) and, in addition, be able to consistently explain the results of our aging experiment that is discussed next.

\begin{figure}
\includegraphics[width=.9\columnwidth]{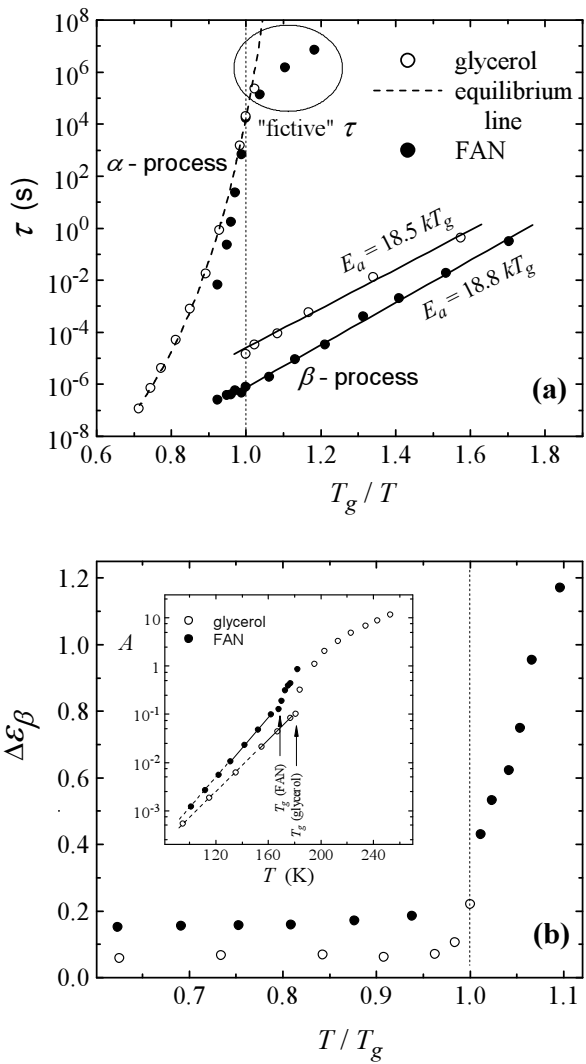}
\caption{\label{parameters} {\bf (a)} Relaxation times of glycerol and FAN in temperature activation plot. {\bf (b)} $\beta$-relaxation strength {\it vs.} reduced temperature; inset: apparent wing amplitude {\it vs.} temperature.
	}
\end{figure}

\begin{figure}
\includegraphics[width=.9\columnwidth]{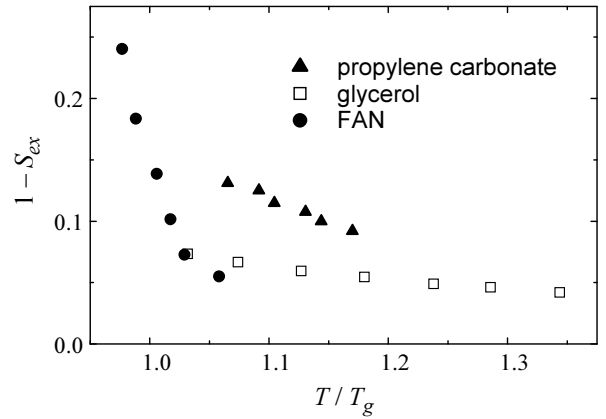}
\caption{\label{strength} Relaxation strength $1-S_{ex}$ of the excess wing (see Eq. \ref{rel_fn}) {\it vs.} reduced temperature.
	}
\end{figure}

Fig. \ref{glycerol}(b) presents results of an aging experiment at 181 K, where the lower spectrum (crosses) was obtained 70 h later than the upper one (open circles). We interpret the difference as solely due to a simultaneous shift of the $\alpha$-peak {\it and} excess wing in the course of aging, with the $\beta$-peak virtually unchanged. This interpretation disagrees with the approach of Ref. \cite{Schneider2000}, where similar effects were explained by an increasing separation of the $\alpha$-peak and a $\beta$-peak during aging. In support of our interpretation, we show in Fig. \ref{glycerol}(b) the same data with the wing contribution subtracted out, assuming for simplicity it is an additive power law $\nu^{-\gamma}$. This results in an $\alpha$-peak flank with the same $\beta =0.63$ as at high temperatures, and a symmetric $\beta$-peak. Note specifically a nearly perfect overlap of the aged and non-aged $\beta$-peaks, which implies that the $\beta$-peak does not change in the course of aging. This analysis yields $\gamma = 0.21$, which is close to the values obtained from the high-temperature unconstrained fits. We thus assume that $\gamma = 0.21$ at {\it all} temperatures and repeat the fits with only {\it three} free parameters left, the relaxation strength of the excess wing $1-S_{ex}$, time constant $\tau_\alpha$, and overall amplitude $\Delta\epsilon$. At 181 and 184 K we allow, in addition, for a $\beta$-contribution, $1-S_\beta >0$. The resulting fits, shown in Fig. \ref{glycerol}(a), are quite satisfactory at all temperatures. 
Included in Fig. \ref{glycerol}(a) are also the individual contributions of the excess wing at several temperatures (dashed lines), which grow with decreasing the temperature, cf. also Fig. \ref{strength}. (Note that the apparent amplitude $A$ of their power-law flanks $A\,\nu ^{-\gamma}$ rather decreases; see also inset in Fig. \ref{parameters}(b).) This is the opposite of the behavior expected for a $\beta$-process. 

The analysis can be extended to temperatures below $T_g$, where the $\alpha$-peak itself is outside of the frequency window. Assuming that the exponent of the excess wing is not changing even below $T_g$, one can isolate the $\beta$-peak and even some residual contribution from the $\alpha$-process by subtracting the (power-law) wing contribution. Some results, shown in Fig. \ref{glycerol}(c), reveal a well resolved $\beta$-peak of glycerol below $T_g$. Its time constant $\tau_\beta$ and amplitude $\Delta \epsilon_\beta$ {\it vs.} temperature are plotted in Fig. \ref{parameters}. Clearly, $\tau_\beta (T)$ follows an Arrhenius behavior. The activation energy of $18.5 \, k T_g$ is similar to that usually found for $\beta$-processes. The amplitude $\Delta \epsilon_\beta$ is virtually temperature independent below $T_g$ and increases above, again as is typical for a $\beta$-process. Regarding the residual contribution from the "frozen" $\alpha$-process (not shown), its "fictive" relaxation time below $T_g$ can be estimated by assuming the same $\alpha$-relaxation strength as in the liquid. The data points then fall below the equilibrium liquid line of $\tau_\alpha (T)$, see Fig. \ref{parameters}(a) (better seen for FAN, which is discussed later), but are expected to converge on it upon sufficiently, even if inaccessibly, long aging. 

Returning to the behavior of the excess wing below $T_g$, its relaxation strength $1-S_{ex}$ is no longer defined. But we can still determine the amplitude $A$ of the associated power law $A\,\nu^{-\gamma}$, see the inset of Fig. \ref{parameters}(b). In contrast to the relaxation strength, the amplitude $A$ increases with temperature. Below $T_g$, the dependence appears to be exponential, $A \propto \exp(T/T_0)$. This is quite similar to our previous results \cite{Kudlik99,Gainaru05}, where we however did not discriminate between the excess wing and a $\beta$-process but interpreted the dielectric spectrum as a nearly constant loss. We therefore extrapolated $A$ exponentially for $T \leq 135$ K, in order to extract $\beta$-contributions at these temperatures, where only a fragment of the peak remains in the frequency window. Above $T_g$, the temperature dependence of $A$ changes, see the inset in Fig. \ref{parameters}(b).

Next we turn to the dielectric spectra of 3-fluoroaniline (FAN) \cite{Kudlik99}, which show a clear $\beta$-process and an excess wing close to $T_g \approx 172$ K, see Fig. \ref{FAN}(a). Again, we determine $\beta$ from the high temperature data, $\beta = 0.52$ at 207 K. As the $\beta$-process is much stronger than in glycerol, the analysis is straightforward; the fits to the complete Eq. (1), with all three contributions, are shown in Fig. \ref{FAN}(a). The exponent of the excess wing was fixed at $\gamma = 0.23$, as obtained from a free fit at 168 K, where the wing is best resolved. No systematic deviation between the fits and data is observed. The $\beta$-peaks, extracted by subtracting the excess wing contribution (see Fig. \ref{FAN}(a)), are shown in Fig. \ref{FAN}(b). At $T \geq 168$ K such decomposition cannot be done, since the additivity assumption becomes too crude; the individual $\beta$-contribution can however be readily extracted from the fit function, see two upper curves in Fig. \ref{FAN}(b). The temperature dependences of the $\beta$-peak's time constant and amplitude appear to be similar to those of glycerol, see Fig. \ref{parameters}. Above $T_g$, the $\beta$-peak shape appears to be independent of the temperature, while $\tau_\beta$ displays a somewhat weaker temperature dependence than at $T<T_g$. Such behavior of a $\beta$-process was first reported by Olsen et al. \cite{Olsen98}. The apparent amplitude $A$ of the excess wing of FAN shows a similar temperature dependence to glycerol, including a change at $T_g$, while its relaxation strength $1-S_{ex}$ is larger and has a stronger temperature dependence than in glycerol, see Fig. \ref{strength}. The difference is well recognized visually, when the spectra of FAN and glycerol are directly compared, see the left-most datasets in Fig. \ref{FAN}(a). Moreover, the relative contribution of the excess wing of FAN can be ignored already above $T=182$ K.

Finally, we analyzed in the same way the data of propylene carbonate ($T_g \approx 158$ K) in the temperature range 163 to 179 K and found $\beta$ = 0.78, $\gamma$ = 0.3. The relaxation strength of the excess wing {\it vs.} temperature is included in Fig. \ref{strength}.

Concluding, we have presented an analysis of the dielectric spectra of three paradigmatic molecular glassformers that leads to a new interpretation of the temperature evolution of the dynamic susceptibility. The long-time part of the main $\alpha$-relaxation keeps its shape, and thus obeys frequency-temperature superposition, at all temperatures. The apparent broadening of the relaxation peak upon cooling is caused solely by an increasing contribution from the excess wing with a temperature-independent exponent. The overall temperature variation, apart from the peak shift, can thus be completely attributed to a variation of the relative weight of the wing. Such a description is only possible by allowing for a $\beta$-process even in cases where none is clearly discernible. For glycerol, this yields a $\beta$-peak with typical properties of the $\beta$-processes, in particular, with virtually the same activation energy as in FAN and similar attempt times. This suggests that characteristic $\beta$-relaxation may indeed be a truly universal phenomenon, as anticipated by Johari and Goldstein \cite{Johari70}. However, and contrary to a widespread interpretation, the excess wing and $\beta$-process appear to be qualitatively different phenomena.

We realize that our results are based on a purely phenomenological spectral shape analysis and, being such, cannot directly help with theoretical understanding of the subject, or explain the physical nature and molecular origin of the processes involved. Yet we believe they do give a new insight that will help to establish a coherent physical picture.

Comparing our current approach with other quantitative phenomenological spectral shape analyses, such as those of Ref. \cite{Blo03,Blo06}, we note that, while our model fits the data just as well and in most cases with less free parameters, the results as to the temperature dependence of the spectral widths are remarkably different. At present, in the absence of a theory and based on purely phenomenological grounds, we find it impossible to judge which one of the approaches along with their respective results is more appropriate.

The authors appreciate valuable discussions with Th. Blochowicz and thank P. Lunkenheimer for kindly supplying dielectric data of glycerol and propylene carbonate. Support through Sonderforschungsbereich 481 is acknowledged.

\end{document}